\documentclass[aps,pra,showpacs,twocolumn,groupedaddress,a4paper]{revtex4}

\usepackage{graphics}

\begin{document}


\title{Breathing mode of rapidly rotating Bose-Einstein condensates}


\author{Gentaro Watanabe$^{a,b}$}

\affiliation{
$^{a}$NORDITA, Blegdamsvej 17, DK-2100 Copenhagen \O, Denmark
\\
$^{b}$The Institute of Chemical and Physical Research (RIKEN),
2-1 Hirosawa, Wako, Saitama 351-0198, Japan
}


\date{\today}

\begin{abstract}
We show that the breathing mode of a rapidly-rotating, 
harmonically-trapped Bose-Einstein condensate may be described 
by a generalized lowest Landau level (LLL) wave function,
in which the oscillator length is treated as a variable.
Using this wave function in a variational Lagrangian formalism, 
we show that the frequency of the breathing mode 
for a two-dimensional cloud is $2\omega_{\perp}$, 
where $\omega_{\perp}$ is the trap frequency.
We also study large-amplitude oscillations and confirm that the above result
is not limited to linear oscillations.
The resulting mode frequency can be understood in terms 
of orbits of a single particle in a harmonic trap.  
The mode frequency is also calculated for a cloud in three dimensions and
the result for the axial breathing mode frequency agrees with recent
experimental data in the rapid rotation regime.
\end{abstract}

\pacs{03.75.Kk, 05.30.Jp, 67.40.Vs, 47.37.+q}

\maketitle


\section{Introduction\label{sect intro}}

The creation of vortices and vortex lattices in Bose-Einstein condensates 
of cold atomic gases
\cite{first,spoon,lattice}
has opened up the study of vortex lattices in a regime 
which cannot be studied in liquid helium 4.  
Due to the diluteness of cold atomic gases, 
it is possible to realize situations in which the vortex core size is 
comparable to the vortex spacing \cite{compara}.

In a seminal work, Ho \cite{ho}
pointed out that the Hamiltonian for a rotating cloud
in a harmonic trap has the same form as that for charged particles
in a uniform magnetic field. 
Based on the similarity between these two systems,
he argued that when the rotation frequency $\Omega$ is close to the transverse
frequency $\omega_{\perp}$ of the trap, almost all particles would condense
into the lowest Landau level (LLL) of the Coriolis force.
Stimulated by this insight, extensive experimental studies have been
performed by the JILA group 
(see, e.g., Coddington {\it et al.} \cite{coddington}),
who have achieved angular velocities in excess of 
$0.99\omega_{\perp}$ at which the cloud contains a vortex lattice
with several hundred vortices \cite{schweikhard}.

The current frontier of the experiments is entering 
the mean-field LLL regime in which $\hbar\Omega$ is comparable to
or larger than the interaction energy, $gn$, leading to a system
whose wave function is dominated by the LLL component. 
Here $n$ is the particle density
and $g\equiv 4\pi\hbar^2 a_{\rm s}/m$ is the two-body interaction strength,
$m$ is the particle mass, and $a_{\rm s}$ is the $s$-wave scattering length.
In this regime, the Gross-Pitaevskii equation
is still applicable because the number of particles $N$
is much larger than that of vortices $N_{\rm v}$.
As shown by ourselves \cite{wbp} and confirmed by 
Cooper {\it et al.} \cite{ckr} (see also Ref.\ \cite{aftalion}), 
a single-particle wave function of the LLL form
is a good approximation to the ground state in the rotating frame, i.e.,
the yrast state, of rapidly rotating condensates.
Even the subtle connection between the vortex lattice distortion 
and the density profile can be described by the simple LLL approximation:
a small distortion of the vortex results in a drastic change
of the density profile from Gaussian to the Thomas-Fermi parabola.

However, limitations of the LLL wave function become apparent
when we consider excited states.
The oscillations of the cloud radius (a so-called breathing mode)
in the direction of the rotation axis 
have been measured in the mean-field LLL regime \cite{schweikhard},
and those in the transverse direction are expected to be measured
in the future experiments.
However, the LLL wave function does not have the flexibility 
to describe these simple collective modes.
In the present work, we consider which degrees of freedom 
in the LLL wave function are relevant to describe the breathing mode and
provide a simple improved treatment, which 
remedies the limitations of the LLL wave function.

In the beginning of the next section, we give the basic formalism
of the present analysis: the variational Lagrangian approach 
\cite{perez_garcia}. 
Using this framework, we then discuss 
the breathing mode of the rapidly rotating
two dimensional Bose-Einstein condensates.
In Section \ref{sect nonlinear},
large-amplitude oscillations are discussed.
In Section \ref{sect interpretation},
we provide a simple physical understanding of the result.
Summary and conclusion are given in Section \ref{conclusion}.
The mode frequency in three dimensions is calculated in Appendix \ref{sect 3d}.

\section{General Formalism\label{sect formalism}}

Let us first see why the standard LLL wave function
cannot describe the breathing mode.
For definiteness, we consider purely two-dimensional motion
in the plane perpendicular to the rotation axis.
We write the condensate wave function $\psi$ in the form
$\psi=\sqrt{N}\phi$, where $\phi$ is normalized such that $|\phi|^2$
integrated over all space is unity.
The LLL wave function can be written as
\begin{equation}
  \phi_{\rm LLL}({\bf r})=A_{\rm LLL}
 \prod_{i=1}^{N_{\rm v}}(\zeta-\zeta_i)\ e^{-r^2/2d_{\perp}^2}\ ,
 \label{LLL}
\end{equation}
where the rotation axis is along the $z$-axis, $\zeta=x+iy$, $\zeta_i$ are
the vortex positions measured in the laboratory frame,
$N_{\rm v}$ is the number of vortices, ${\bf r}=(x, y)$, and
$d_{\perp}=\sqrt{\hbar/m\omega_{\perp}}$ is the
transverse oscillator length.

Employing Eq.\ (\ref{LLL}), we find the angular momentum per particle 
to be given by
\begin{equation}
  L_z \simeq \hbar \int d^2r\, \left(\frac{r^2}{d_{\perp}^2}-1\right)\ 
\langle |\phi_{\rm LLL}|^2 \rangle
= \hbar \left(\frac{\langle r^2\rangle}{d_{\perp}^2}-1\right)\ ,
\label{lzLLL}
\end{equation}
where $\langle |\phi_{\rm LLL}|^2 \rangle$ is the density profile
smoothed over an area of linear size large compared to the vortex separation
but smaller than the radial extent of the cloud.
For a uniform density of vortices, the density profile has a Gaussian form,
\begin{equation}
  \langle |\phi_{\rm LLL}^{\rm G}|^2 \rangle =\nu(r)
  = \frac{1}{\pi \sigma^2} e^{-r^2/\sigma^2}\ ,
\label{gaussian}
\end{equation}
where $\sigma$ is the width parameter 
given by the vortex density $n_{\rm v}$ as 
$n_{\rm v}=1/(\pi d_\perp^2) - 1/(\pi\sigma^2)$ \cite{ho}.
In this case one finds 
\begin{equation}
  L_z \simeq \hbar \left(\frac{\sigma^2}{d_{\perp}^2}-1\right)\ .
\label{lzLLLgauss}
\end{equation}
In the real situation, interactions distort the lattice and the density profile
becomes a Thomas-Fermi parabola \cite{wbp,ckr,aftalion},
\begin{equation}
  \langle |\phi_{\rm LLL}^{\rm TF}|^2 \rangle =\nu(r)
  = \nu(0) \left(1-\frac{r^2}{R_{\perp}^2}\right)\ ,
\label{tf}
\end{equation}
with $\nu(0)=2/\pi R_{\perp}^2$, where $R_{\perp}$ is the cloud radius.
In this case the angular momentum is
\begin{equation}
  L_z \simeq \hbar \left(\frac{R_{\perp}^2}{3 d_{\perp}^2} - 1\right)\ .
\label{lzLLLtf}
\end{equation}
Here the superscripts ``G'' and ``TF'' denote that the density profile is
Gaussian and Thomas-Fermi, respectively.

Equations (\ref{lzLLLgauss}) and (\ref{lzLLLtf}) show that, 
for the LLL wave function,
the cloud width parameter $\sigma$ or the cloud radius $R_{\perp}$ is fixed
if the angular momentum is constant, i.e.,
the breathing mode cannot be described by the standard LLL wave function.

\subsection{Extended Lowest Landau Level Wave Function\label{sect Ex}}

To describe the breathing mode in the rapidly rotating limit, 
we introduce an extended lowest Landau level wave function,
which keeps the LLL form but allows the oscillator length 
of the trap (i.e., the trap frequency) to be adjustable:
\begin{equation}
  \phi_{\rm ex}({\bf r})=A_{\rm ex}
 \prod_{i=1}^{N_{\rm v}}(\zeta-\zeta_i)\ 
  \exp{\left[-\left(\frac{1}{\lambda^2}-i\beta\right)\,
  \frac{r^2}{2d_{\perp}^2}\right]}\ .
 \label{Ex}
\end{equation}
The dynamical variable $\lambda$ describes 
the variation of the particle density.
The wave function (\ref{Ex}), like the original LLL wave function (\ref{LLL}), 
has a Gaussian smoothed density profile if the lattice is uniform, 
but with a modified width parameter given by 
\begin{equation}
  n_{\rm v}=\frac{1}{\pi \lambda^2 d_\perp^2}-\frac{1}{\pi\sigma^2}\ .
\end{equation}
The other dynamical variable $\beta$ generates a radial velocity field $v_r$
which causes a homologous change of the density profile \cite{castin}
and the corresponding velocity is
\begin{equation}
  v_{r}=\frac{\hbar}{m}\ \frac{\partial}{\partial r}
  \left(\frac{\beta r^2}{2d_{\perp}^2}\right)
  =\frac{\hbar}{m}\frac{\beta}{d_{\perp}^2}\ r\ .
\end{equation}

In the present analysis, we use the variational Lagrangian formalism
\cite{perez_garcia}.
The Lagrangian per particle consists of the time-dependent part $T$ and 
the energy functional $E$:
\begin{equation}
  {\cal L}[\phi_{\rm ex}]
  =T[\phi_{\rm ex}]-E[\phi_{\rm ex}]\ ,
\end{equation}
with
\begin{equation}
  E[\phi_{\rm ex}]
  =K[\phi_{\rm ex}]+V[\phi_{\rm ex}]+E_{\rm int}[\phi_{\rm ex}]\ ,
\end{equation}
where $K$, $V$, and $E_{\rm int}$ are the kinetic, potential, and interaction
energies.
Each term in the Lagrangian functional for the extended LLL wave function 
can be written as
\begin{widetext}
\begin{eqnarray}
  T[\phi_{\rm ex}]
  &=& \frac{i\hbar}{2} \int d^2r \left[\phi_{\rm ex}^* 
  \frac{\partial \phi_{\rm ex}}{\partial t}
  - \phi_{\rm ex} \frac{\partial \phi_{\rm ex}^*}{\partial t}\right]
  \nonumber\\
  &=& \frac{\hbar}{2} \int d^2r \left[
  \sum_{i=1}^{N_{\rm v}} \frac{2({\bf r}-{\bf r}_i)\times \dot{{\bf r}}_i}
  {|{\bf r}-{\bf r}_i|^2} - \frac{\dot{\beta} r^2}{d_{\perp}^2}\right]
  |\phi_{\rm ex}|^2\ ,\label{td}\\
  K[\phi_{\rm ex}]
  &=& -\frac{\hbar^2}{2m} \int d^2r\ 
  \phi_{\rm ex}^* \nabla_{\perp}^2 \phi_{\rm ex}
  = -\frac{2\hbar^2}{m} \int d^2r\
  \phi_{\rm ex}^* \partial_{\zeta^*}\partial_{\zeta} \phi_{\rm ex}
  \nonumber\\
  &=& \frac{\hbar\omega_{\perp}}{2} \int d^2r 
  \left(\frac{1}{\lambda^4}+\beta^2\right)\frac{r^2}{d_{\perp}^2}
  |\phi_{\rm ex}|^2\ ,\label{kin}\\
  V[\phi_{\rm ex}] &=& \frac{m\omega_{\perp}^2}{2} \int d^2r\ 
  r^2 |\phi_{\rm ex}|^2\ ,\label{pot}\\
\mbox{and}\qquad\qquad\qquad&&\nonumber\\
  E_{\rm int}[\phi_{\rm ex}] 
  &=& \frac{g_{\rm 2D}}{2} \int d^2r\ |\phi_{\rm ex}|^4\ ,\label{int}
\end{eqnarray}
\end{widetext}
where $g_{\rm 2D}$ is the effective interaction parameter in two dimensions. 
If the system is uniform in the $z$-direction, 
$g_{\rm 2D}=Ng/Z$, where $Z$ is the axial extent of the cloud, 
while if the system in the $z$-direction is in the
ground state of a particle in a harmonic potential of
frequency $\omega_z$, then $g_{\rm 2D}=Ng/(d_z\sqrt{2\pi})$, 
where $d_z=\sqrt{\hbar/m\omega_z}$ is the oscillator length
in the $z$-direction \cite{fetter2}.
The angular momentum per particle for $\phi_{\rm ex}$ is
\begin{eqnarray}
  L_z[\phi_{\rm ex}] &=& \int d^2r\ 
  \phi_{\rm ex}^* \left[{\bf r}\times(-i\hbar\nabla_{\perp})\right]_z 
  \phi_{\rm ex}
  \nonumber\\
  &=& \hbar \int d^2r\
  \phi_{\rm ex}^* (\zeta\partial_{\zeta}-\zeta^*\partial_{\zeta^*})
  \phi_{\rm ex}
  \nonumber\\
  &=& \hbar \int d^2r 
  \left(\frac{r^2}{\lambda^2 d_{\perp}^2}-1\right)
  |\phi_{\rm ex}|^2\ .\label{lz}
\end{eqnarray}

Let us now consider the first term of Eq.\ (\ref{td}),
which we denote by $T_1$.
The velocity of the vortex $\dot{{\bf r}}_i$ in the laboratory frame
has the azimuthal component $v_{\phi,i}\simeq r_i\Omega$ due to the bulk
rotation and the radial component $v_{r,i}\simeq \delta R~ \omega r_i/R$,
where $R$ is the scale of the radial extent of the cloud,
$\delta R$ and $\omega$ are the amplitude and the frequency
of the breathing oscillation.
Employing the averaged vortex approximation as we will do in the later 
discussion, only the azimuthal velocity gives a nonzero contribution
because the coarse-grained density profile is axisymmetric:
\begin{equation}
T_1=\hbar \sum_{i=1}^{N_{\rm v}} \int d^2r\
   \frac{({\bf r}-{\bf r}_i)\times \hat{\mbox{\boldmath $\phi$}_i} v_{\phi,i}}
  {|{\bf r}-{\bf r}_i|^2} \nu(r)\ ,
\end{equation}
where $\hat{\mbox{\boldmath $\phi$}_i}$ is the unit vector of the azimuthal
direction measured from vortex $i$.  
We can readily see that the integral of $T_1$ vanishes 
when the density is uniform.
For the inhomogeneous cloud, the dominant contribution of $T_1$
comes from vortices at $r_i\sim R$; vortices at $r_i \ll R$ give
only small contribution because the cancellation of the cross product
between different ${\bf r}$ is efficient \cite{note lattice}.
The integral of $T_1$ for each vortex at $r_i\sim R$ yields
$\sim \int d^2r\ \Omega \nu(r)=\Omega$ because $|{\bf r}-{\bf r}_i|\sim R$
and $v_{\phi,i}\sim R\Omega$.
The number of vortices in the surface region is proportional to
$n_{\rm v}\ell R \sim N_{\rm v} \ell/R$,
where $\pi\ell^2$ gives the area per vortex.
Thus, for a cloud with a large vortex lattice, $T_1$ scales as
\begin{equation}
  T_1\propto \hbar\Omega N_{\rm v} \frac{\ell}{R}\ ,
\end{equation}
which becomes negligible compared to the second term of Eq.\ (\ref{td})
$T_2\sim \hbar\omega N_{\rm v}$ if $R \gg \ell$, i.e., $N_{\rm v}^{1/2}\gg1$
[we will see later that $\omega \sim \Omega (\simeq \omega_{\perp})$].
In the later discussion, we assume that the vortex lattice is large enough
that the effect of $T_1$ may be neglected.

\subsection{Gaussian Profile\label{sect uniform}}

Let us first consider a cloud with a uniform vortex lattice.
Using the averaged vortex approximation, we replace
$|\phi_{\rm ex}|^2$ in Eqs.\ (\ref{td}) - (\ref{int})
by the smoothed density profile of a cloud (\ref{gaussian})
with a uniform vortex lattice.  We then obtain
\begin{eqnarray}
  T[\phi_{\rm ex}^{\rm G}] 
  &\simeq& -\frac{\hbar}{2} \frac{\dot{\beta}\sigma^2}{d_{\perp}^2}\ ,\\
  K[\phi_{\rm ex}^{\rm G}]
  &\simeq& \frac{\hbar\omega_{\perp}}{2}
  \left(\frac{1}{\lambda^4}+\beta^2\right)\frac{\sigma^2}{d_{\perp}^2}\ ,\\
  V[\phi_{\rm ex}^{\rm G}]
  &\simeq& \frac{\hbar\omega_{\perp}}{2} \frac{\sigma^2}{d_{\perp}^2}\ ,\\
  E_{\rm int}[\phi_{\rm ex}^{\rm G}] 
  &\simeq& \frac{b g_{\rm 2D}}{4\pi\sigma^2}\ ,
\end{eqnarray}
where $b\equiv\langle |\phi_{\rm ex}|^4 \rangle/
\langle |\phi_{\rm ex}|^2 \rangle^2$ 
is the Abrikosov parameter, which is comparable to unity.

Similarly, Eq.\ (\ref{lz}) reads
\begin{equation}
  L_z[\phi_{\rm ex}^{\rm G}]
  \simeq \hbar\left(\frac{\sigma^2}{\lambda^2 d_{\perp}^2}-1\right)\ .
\label{lz gauss}
\end{equation}
Due to the angular momentum conservation,
the dynamical variables $\sigma$ and $\lambda$ are not independent
by Eq.\ (\ref{lz gauss}).
Writing the constant value of $L_z$ as $l_z$,
$\lambda$ is expressed as
\begin{eqnarray}
  \frac{1}{\lambda^2}=\frac{d_{\perp}^2}{\sigma^2}
  \left(\frac{l_z}{\hbar}+1\right)
  \equiv \frac{d_{\perp}^2}{\sigma^2} l\ .
\label{lambda}
\end{eqnarray}
Using Eq.\ (\ref{lambda}) to eliminate $\lambda$ from the Lagrangian, we obtain
\begin{widetext}
\begin{eqnarray}
  {\cal L}[\phi_{\rm ex}^{\rm G}]
  =-\frac{\hbar}{2} \frac{\dot{\beta}\sigma^2}{d_{\perp}^2}
  -\left[\frac{\hbar\omega_{\perp}}{2} 
  \left\{ l^2\frac{d_{\perp}^2}{\sigma^2} 
  + (\beta^2+1)\frac{\sigma^2}{d_{\perp}^2} \right\}
  +\frac{b g_{\rm 2D}}{4\pi\sigma^2}\right]\ .
\label{lagrangian_gauss}
\end{eqnarray}
\end{widetext}

Thus the Euler-Lagrange equations 
for the dynamical variables $\beta$ and $\sigma$
\begin{eqnarray}
  \frac{d}{dt}\left(\frac{\partial {\cal L[\phi_{\rm ex}^{\rm G}]}}{\partial \dot{\beta}}\right)
  - \frac{\partial {\cal L[\phi_{\rm ex}^{\rm G}]}}{\partial \beta}&=&0\ ,\\
  \frac{d}{dt}\left(\frac{\partial {\cal L[\phi_{\rm ex}^{\rm G}]}}{\partial \dot{\sigma}}\right)
  - \frac{\partial {\cal L[\phi_{\rm ex}^{\rm G}]}}{\partial \sigma}&=&0\ ,
\end{eqnarray}
lead to the following equations,
\begin{equation}
  \dot{\sigma} = \omega_{\perp} \beta \sigma\ ,
\end{equation}
and
\begin{equation}
  \left(\frac{\dot{\beta}}{\omega_{\perp}}+\beta^2+1\right)
  \frac{\sigma^4}{d_{\perp}^4} - l^2 - \kappa = 0\ ,
\end{equation}
respectively;
where $\kappa\equiv mbg_{\rm 2D}/(2\pi\hbar^2)$ is the dimensionless
interaction strength.
Combining the above two equations, we obtain an equation of motion
of the cloud width parameter $\sigma$:
\begin{equation}
  \frac{1}{\omega_{\perp}^2 d_{\perp}^4} \sigma^3\ \ddot{\sigma}
  + \frac{\sigma^4}{d_{\perp}^4} - l^2 - \kappa = 0\ .
\label{eom gauss}
\end{equation}
According to Eq.\ (\ref{eom gauss}), the cloud width parameter $\sigma_0$
in the equilibrium state is given as
\begin{equation}
  \sigma_0^4 = d_{\perp}^4(l^2+\kappa)\ .
\label{sigma0}
\end{equation}
We note that this expression coincides with the width parameter 
of the minimum energy state 
obtained by a variational calculation in the rotating frame.
Now we suppose $\sigma$ oscillates around $\sigma_0$ as
\begin{equation}
  \sigma(t)=\sigma_0 + \delta\sigma(t)\ ,
\end{equation}
and its deviation $\delta\sigma(t)$ is small.
Linearizing Eq.\ (\ref{eom gauss}) with respect to $\delta\sigma$ yields
\begin{equation}
  \delta\ddot{\sigma}(t) + (2\omega_{\perp})^2 \delta\sigma(t)=0\ .
\label{linearized eom gauss}
\end{equation}
Thus the breathing mode frequency $\omega$ in the LLL regime 
for a Gaussian profile is
\begin{equation}
  \omega=2\omega_{\perp}\ .
\end{equation}
This result agrees with what one obtains from calculations based on
hydrodynamics if one takes the limit of a trapping potential that
is independent of $z$ \cite{sedrakian,cozzini}.  In the case of
Ref.\ \cite{sedrakian}, the polytropic index must be put equal to the
value two appropriate for a dilute Bose gas.

It is notable that the interaction energy scales in the same way
as the rotational kinetic energy (the $l^2$ term)
in the Lagrangian (\ref{lagrangian_gauss}) and in the equation of motion
(\ref{eom gauss}) \cite{note scaling}.
Thus the interaction energy enters only in the combination $l^2+\kappa$, so 
it does not affect the frequency of the breathing oscillation 
\cite{note nonlinear}.
This is a remarkable feature of the two-dimensional system \cite{note scaling}.

In the case of a three-dimensional rotating cloud
trapped in a potential $V=m\omega_{\perp}(x^2+y^2)/2+m\omega_z z^2/2$,
the interaction energy is given as $E_{\rm int}\sim 1/(\sigma^2 R_z)$, where
$R_z$ is the radius (width parameter) of the cloud in the $z$-direction
when the density profile of this direction is the Thomas-Fermi parabola 
(Gaussian).  Thus, for a non-zero interaction strength, the two-dimensional
monopole oscillation in the $xy$-plane couples with the one-dimensional
monopole oscillation along the $z$-axis whose frequency is $\sqrt{3} \omega_z$
(in the case where the zero-point energy of the $z$-direction is negligible
compared to the interaction energy)
and the resulting mode frequency is modified from $2\omega_{\perp}$
(see Appendix \ref{sect 3d}).

\subsection{Thomas-Fermi Profile\label{sect distorted}}

As in the previous section, we adopt the averaged vortex approximation.
For the inverted parabolic density profile of Eq.\ (\ref{tf})
for a distorted vortex lattice, Eqs.\ (\ref{td}) - (\ref{int})
can be written as
\begin{eqnarray}
  T[\phi_{\rm ex}^{\rm TF}] 
  &\simeq& -\frac{\hbar}{2} \frac{\dot{\beta}R_{\perp}^2}{3d_{\perp}^2}\ ,\\
  K[\phi_{\rm ex}^{\rm TF}]
  &\simeq& \frac{\hbar\omega_{\perp}}{2}
  \left(\frac{1}{\lambda^4}+\beta^2\right)\frac{R_{\perp}^2}{3d_{\perp}^2}\ ,\\
  V[\phi_{\rm ex}^{\rm TF}]
  &\simeq& \frac{\hbar\omega_{\perp}}{2}
  \frac{R_{\perp}^2}{3d_{\perp}^2}\ ,\\
  E_{\rm int}[\phi_{\rm ex}^{\rm TF}]
  &\simeq& \frac{2 b g_{\rm 2D}}{3\pi R_{\perp}^2}\ .
\end{eqnarray}
Using the angular momentum conservation:
\begin{eqnarray}
  l_z\equiv L_z[\phi_{\rm ex}^{\rm TF}]
  \simeq \hbar\left(\frac{R_{\perp}^2}{3\lambda^2 d_{\perp}^2}-1\right)\ ,
\label{lztf}
\end{eqnarray}
$\lambda$ can be expressed as
\begin{eqnarray}
  \frac{1}{\lambda^2}=3\frac{d_{\perp}^2}{R_{\perp}^2}
  \left(\frac{l_z}{\hbar}+1\right)
  \equiv 3\frac{d_{\perp}^2}{R_{\perp}^2} l\ .
\end{eqnarray}
Thus the Lagrangian can be written with only $R_{\perp}$ and $\beta$:
\begin{widetext}
\begin{eqnarray}
  {\cal L}[\phi_{\rm ex}^{\rm TF}]
  =-\frac{\hbar}{2} \frac{\dot{\beta}R_{\perp}^2}{3d_{\perp}^2}
  -\left[\frac{\hbar\omega_{\perp}}{2} 
  \left\{l^2\frac{3d_{\perp}^2}{R_{\perp}^2} 
  + (\beta^2+1)\frac{R_{\perp}^2}{3d_{\perp}^2} \right\}
  +\frac{2 b g_{\rm 2D}}{3\pi R_{\perp}^2}\right]\ .
\label{lagrangian_tf}
\end{eqnarray}
\end{widetext}

The Euler-Lagrange equation for $\beta$ leads to
\begin{equation}
  \dot{R}_{\perp} = \omega_{\perp} \beta R_{\perp}\ ,
\label{euler beta tf}
\end{equation}
and that for $R_{\perp}$ to
\begin{equation}
  \left(\frac{\dot{\beta}}{\omega_{\perp}}+\beta^2+1\right)
  \frac{R_{\perp}^4}{d_{\perp}^4} - 9\ l^2 - 8\kappa =0\ .
\label{euler r tf}
\end{equation}
Combining Eqs.\ (\ref{euler beta tf}) and (\ref{euler r tf}), we obtain
\begin{equation}
  \frac{1}{\omega_{\perp}^2 d_{\perp}^4} R_{\perp}^3\ \ddot{R}_{\perp}
  + \frac{R_{\perp}^4}{d_{\perp}^4} - 9\ l^2 - 8\kappa = 0\ .
\label{eom tf}
\end{equation}
The cloud radius $R_{\perp,0}$ in the equilibrium state is thus given as
\begin{equation}
R_{\perp,0}^4=d_{\perp}^4 (9 l^2+8\kappa)\ ,
\label{R0}
\end{equation}
which coincides with the cloud radius
obtained by minimizing the energy in the rotating frame.
Again we notice that the interaction energy scales in the same way
as the rotational kinetic energy, and it does not affect 
the breathing mode frequency \cite{note nonlinear}
(see the Appendix for the three-dimensional case).

A linearized equation of motion for a small oscillation of
$R_{\perp}(t)=R_{\perp, 0} + \delta R_{\perp}(t)$ is
\begin{equation}
  \delta\ddot{R}_{\perp}(t) + (2\omega_{\perp})^2\ \delta R_{\perp}(t) =0\ .
\label{linearized eom tf}
\end{equation}
According to Eq.\ (\ref{linearized eom tf}), 
the breathing mode frequency $\omega$
in the LLL regime for a cloud with a Thomas-Fermi profile is also
\begin{equation}
  \omega=2\omega_{\perp}\ ,
\end{equation}
which coincides with results in the slow rotation regime 
\cite{sedrakian,cozzini}.
It is also noted that this result is in accord with measurements
\cite{compara} in the slowly rotating regime.

\section{Non-linear Oscillations
\label{sect nonlinear}}

In the present section, we consider large-amplitude oscillations.
However, we should mention that the several assumptions in the
preceding discussion can break down in the non-linear regime
even if the mean-field Gross-Pitaevskii theory is still a good approximation,
and thus the validity of the result in the present section is limited.
First of all, for the extended LLL wave function to be valid,
the density should be always low enough to ensure 
that the interaction energy is 
much smaller than the energy gap between the LLL and higher Landau levels,
i.e., $gn\ll\hbar\omega_{\perp}$ or $Na_{\rm s}/Z \ll R^2/d_{\perp}^2$,
where $R$ denotes the cloud width or radius.
To keep the two-dimensional character of the system, we require
$\hbar\omega_z\gg m\dot{R}^2$;
otherwise, oscillations in the $z$-direction are excited.
One should also note the criterion for the first term $T_1$ 
in the time-dependent part of the Lagrangian (\ref{td}) to be negligible
compared to the second term $T_2$.
If the amplitude is large, the cloud can shrink significantly
as the angular velocity (and $T_1$) can become large.
Feynman's relation gives the angular velocity of the cloud as
$\Omega\simeq\hbar/(m\ell^2) \sim \hbar N_{\rm v}/(mR^2)$; then we have
$T_1\sim\hbar\Omega N_{\rm v}^{1/2} \sim \hbar^2 N_{\rm v}^{3/2}/(mR^2)$.
Thus the criterion for $T_1\ll T_2 \sim \hbar\omega_{\perp}N_{\rm v}$ 
leads to $R\gg N_{\rm v}^{1/4}d_{\perp}$.

\begin{figure}
\begin{center}\vspace{0.0cm}
\rotatebox{0}{
\resizebox{7.0cm}{!}
{\includegraphics{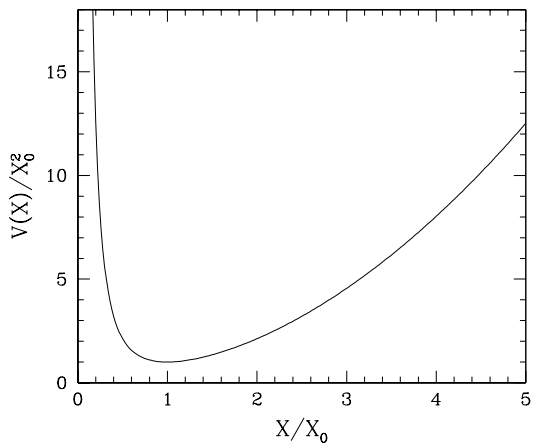}}}
\caption{\label{fig pot}
  Effective potential $V(X)=X^2/2+X_0^4/2X^2$ of the motion 
  of the radial extent of the cloud $X$.
  }
\end{center}
\end{figure}

The equations of motion (\ref{eom gauss}) and (\ref{eom tf})
for both the Gaussian and Thomas-Fermi profiles can be written
as the following general form:
\begin{equation}
\frac{d^2 X}{dt^2}+\frac{d}{dX}\left(\frac{X^2}{2}+\frac{X_0^4}{2X^2}\right)
=0\ ,
\label{eom general}
\end{equation}
where $X$ denotes $\sigma$ or $R_{\perp}$; $X_0$ corresponds to $\sigma_0$
or $R_{\perp,0}$ given by Eqs.\ (\ref{sigma0}) and (\ref{R0}), respectively.
Here we measure the length and time in units of 
$d_{\perp}$ and $\omega_{\perp}^{-1}$.
Equation (\ref{eom general}) has the same form as the Newton's 
equation of motion for a particle moving in a potential 
$V(X)\equiv X^2/2+X_0^4/2X^2$ (see Fig.\ \ref{fig pot}).
This potential shows the restoring force caused by the first term
when the cloud expands and the strong centrifugal repulsion by the second term
when the cloud contracts.

Suppose that, at $t=0$, the cloud has its equilibrium size $X=X_0$ but
with velocity $\dot{X}(0)=\dot{X}_0$ 
to excite a breathing oscillation.
Thus the initial kinetic and potential energies are $\dot{X}_0^2/2$
and $V(X_0)=X_0^2$, respectively.
The oscillation enters the non-linear regime 
when $\dot{X}_0^2/2 \agt X_0^2$ because the potential is no longer
approximated by the harmonic one.
We thus introduce the ``non-linearity parameter'' $\chi$ as
\begin{equation}
  \chi \equiv \frac{\dot{X}_0^2}{2X_0^2}\ ;
\end{equation}
$\chi\ll 1$ in the linear regime and $\chi \agt 1$ in the non-linear regime.

Multiplying Eq.\ (\ref{eom general}) by $\dot{X}$ and writing as
$\dot{X}\ddot{X}=(1/2) (d\dot{X}^2/dt)$ and $\dot{X} (dV/dX)= dV/dt$,
we can integrate this equation analytically and finally obtain
\begin{eqnarray}
  t-t_0 &=& \pm \int \frac{dX}{\sqrt{2\{E-V(X)\}}}\nonumber\\
  &=& \mp \frac{1}{2} \arctan{\left[\frac{E-X^2}{X\sqrt{2\{E-V(X)\}}}\right]}\ ,
\end{eqnarray}
or
\begin{equation}
  \tan{[\mp 2(t-t_0)]} = \frac{E-X^2}{X\sqrt{2\{E-V(X)\}}}\ ,
\label{tant}
\end{equation}
where $t_0$ and $E\equiv\dot{X}_0^2/2+V(X_0)=\dot{X}_0^2/2+X_0^2$ 
are the integration constants.
We can see from Eq.\ (\ref{tant}) that the period $\tau$ of the oscillation
is $\tau=\pi$ (in units of $\omega_{\perp}^{-1}$) and thus the mode frequency
is $2\pi/\tau=2\pi/(\pi\omega_{\perp}^{-1})=2\omega_{\perp}$
even in the non-linear regime.

\begin{figure}
\begin{center}\vspace{0.0cm}
\rotatebox{0}{
\resizebox{8.2cm}{!}
{\includegraphics{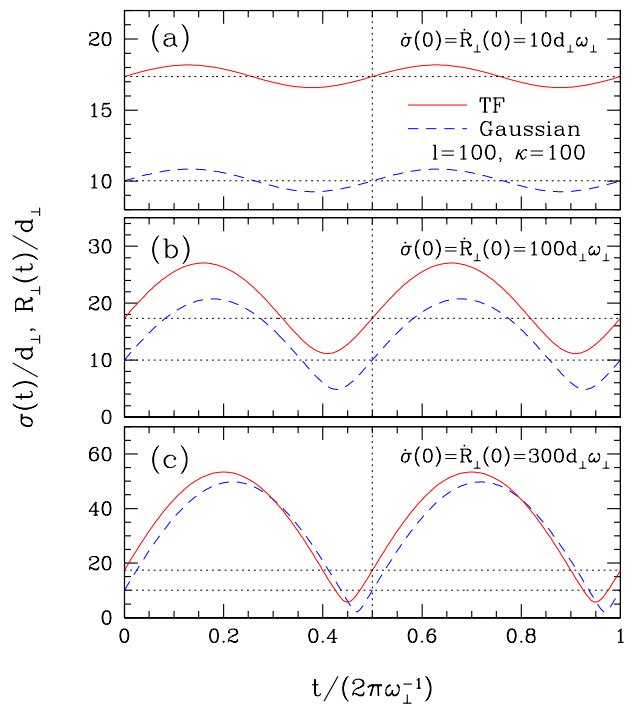}}}
\caption{\label{fig nonlin_osc} (Color online)\quad
  Time evolution of the width parameter $\sigma$ 
  and the radius $R_{\perp}$ of the cloud
  from the linear (a) to the strongly non-linear (c) oscillations.
  }
\end{center}
\end{figure}

In Fig.\ \ref{fig nonlin_osc}, we show the time evolution of
the width parameter $\sigma$ and the radius $R_{\perp}$ of the cloud
for some initial conditions.
In this plot, we set $l=100$ and $\kappa=100$, which are values
appropriate for recent experiments \cite{schweikhard}.
The radial extent of the equilibrium state corresponding to these values
of $l$ and $\kappa$ is $\sigma_0\simeq 10.02d_{\perp}$ for the Gaussian profile
and $R_{\perp,0}\simeq 17.36d_{\perp}$ for the Thomas-Fermi one.
Figure \ref{fig nonlin_osc}(a) shows linear oscillations with $\chi<1$
and the amplitudes of $X$ are almost symmetric above and below $X_0$
in the both cases.
Figure \ref{fig nonlin_osc}(b) shows a case
in the non-linear regime with $\chi=O(10)$.
We observe that the oscillations are asymmetric and this feature is more
prominent in the Gaussian case than in the Thomas-Fermi one.
In Fig.\ \ref{fig nonlin_osc}(c), we show strongly non-linear oscillations
with $\chi>100$ as a demonstration (however, $T_1$ is no longer negligible
in this case).
We can see that the mode frequency $\omega=2\omega_{\perp}$
even in the strongly non-linear regime.

\section{Physical Interpretation of $\omega=2\omega_{\perp}$
\label{sect interpretation}}

In the preceding discussion, we have seen that, in the rapidly rotating regime,
the frequency of the breathing mode of a two dimensional cloud is 
$\omega=2\omega_{\perp}$
as in the slow rotation regime \cite{sedrakian,cozzini,choi,mizushima}
(and in the non-rotating case \cite{pr}).
Furthermore, this result is not limited to small amplitude oscillations.
We can understand the above robustness of the result $\omega=2\omega_{\perp}$ 
for the two dimensional rotating cloud
in the simplest way by focusing on the orbit of a single particle.
Suppose a particle of mass $m$ moves in a circular orbit of radius $r$.
The frequency of oscillations when the circular particle orbit is perturbed
can be interpreted as the breathing mode frequency of the cloud.
Now we assume that each particle moves in accord with
the cloud motion, i.e., $a\equiv r/X$ is constant,
where $X$ denotes $\sigma$ or $R_{\perp}$
in the case of the Gaussian profile or the Thomas-Fermi one, respectively.
In terms of the variable $a$, the density of the cloud can be written as
$n(r)=n(0)\exp{(-r^2/\sigma^2)}=Na^2\exp{(-a^2)}/(\pi r^2)$ 
for the Gaussian profile and
$n(r)=n(0)(1-r^2/R_{\perp}^2)=2Na^2(1-a^2)/(\pi r^2)$
for the Thomas-Fermi one.
The interaction energy of a single particle is given by
$g_{\rm 2D}n(r)=\gamma/r^2$, where
$\gamma\equiv (Ng_{\rm 2D}a^2/\pi)\exp{(-a^2)}$ in the Gaussian case and
$\gamma\equiv (2Ng_{\rm 2D}a^2/\pi) (1-a^2)$ in the Thomas-Fermi case.

\begin{figure}
\begin{center}\vspace{0.0cm}
\rotatebox{0}{
\resizebox{8.5cm}{!}
{\includegraphics{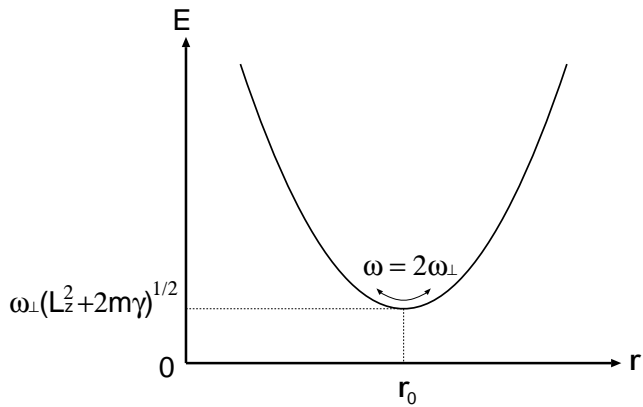}}}
\caption{\label{fig phys mean}
  Schematic picture of the energy of a particle rotating in a harmonic trap
  of the frequency $\omega_{\perp}$.
  Oscillation frequency of the radius of the orbit is $2\omega_{\perp}$.
  }
\end{center}
\end{figure}

The energy of the particle is
\begin{equation}
  E=\frac{mv^2}{2}+\frac{m\omega_{\perp}^2}{2}r^2 +\frac{\gamma}{r^2}
  =\frac{L_z^2}{2mr^2}+\frac{m\omega_{\perp}^2}{2}r^2 +\frac{\gamma}{r^2}\ ,
\end{equation}
where $v$ is the azimuthal velocity and 
$L_z = r m v$ is the angular momentum of the particle,
which is a constant of motion.
The radius $r_0$ of the unperturbed orbit is given by
$\left.\partial E/\partial r\right|_{r=r_0}
=-L_z^2/(mr_0^3)+m\omega_{\perp}^2 r_0 -2\gamma/r_0^3=0$ :
\begin{equation}
  r_0^2=\frac{(L_z^2+2m\gamma)^{1/2}}{m\omega_{\perp}}\ .
\end{equation}
We then give a small perturbation $\delta r$ of $r$ and obtain
\begin{eqnarray}
  E&\simeq&\frac{L_z^2}{2mr_0^2}+\frac{m\omega_{\perp}^2}{2}r_0^2
  + \frac{\gamma}{r_0^2}
  +\frac{m(2\omega_{\perp})^2}{2} \delta r^2\nonumber\\
  &=&\omega_{\perp} (L_z^2+2m\gamma)^{1/2}
  +\frac{m(2\omega_{\perp})^2}{2} \delta r^2\ ,
\end{eqnarray}
which is shown schematically in Fig.\ \ref{fig phys mean}.
The first term gives the energy of the unperturbed motion;
the second term shows that the frequency of the radial oscillation
equals to $2\omega_{\perp}$.
Note that the interaction energy, which has the same $r^{-2}$ dependence
as the rotational kinetic energy, just gives a correction
to the equilibrium state and it does not affect the dynamics.
Thus the breathing mode frequency directly reflects
the general property of the rotating single particle motion in two dimensions.

\section{Summary and Conclusion\label{conclusion}}

In this work, we have investigated the breathing oscillation of a
rapidly rotating two-dimensional Bose-Einstein condensate.
Using the variational extended LLL wave function,
which incorporates the change of the vortex density,
we have shown that the mode frequency is $\omega=2\omega_{\perp}$,
as in the slow rotation regime.
It would be valuable to confirm this prediction
in future experiments
on rapidly rotating Bose-Einstein condensates.
There we have seen that the modulation of the oscillator length 
in the original LLL wave function is an essential degree of freedom 
to describe the breathing mode.
We have also studied large-amplitude oscillations;
we have observed that the amplitude becomes asymmetric 
due to the non-linear effect, but the frequency is still $2\omega_{\perp}$.
Finally, we have provided a simple and physical understanding
of the result $\omega=2\omega_{\perp}$ and its robustness
for two-dimensional rotating clouds.

\begin{acknowledgements}
The author is grateful to C. J. Pethick for helpful discussion, comments,
and support to complete this work.
He also thanks V. Schweikhard for a valuable conversation and
H. M. Nielsen and L. M. Jensen for helpful discussions.
This work was supported in part by the Nishina Memorial Foundation,
and by the JSPS Postdoctoral Fellowship for Research Abroad.
\end{acknowledgements}

\appendix
\section{Breathing Mode in three dimensions\label{sect 3d}}

We consider the breathing mode in the case of three dimensions
and see how the mode frequency is modified from $2\omega_{\perp}$.
In the following analysis, we assume that the radial density profile
is the Thomas-Fermi parabola and the axial one is Gaussian.
Thus the extended LLL wave function in this case is
\begin{eqnarray}
  \phi_{\rm ex}({\bf r})&=&A_{\rm ex}
 \prod_{i=1}^{N_{\rm v}}(\zeta-\zeta_i)\ 
  \exp{\left[-\left(\frac{1}{\lambda^2}-i\beta\right)\,
  \frac{r^2}{2d_{\perp}^2}\right]}\nonumber\\
&&\times\exp{\left[-\frac{z^2}{2\sigma_z^2}+i\gamma\frac{z^2}{2d_z^2}\right]}\ ,
 \label{Ex3d}
\end{eqnarray}
where $\sigma_z$ is the width parameter of the axial density profile 
and $\gamma$ is the dynamical variable describing the velocity field, 
which causes the homologous change of the axial density profile.
The coarse-grained density profile can be written as
\begin{equation}
  \langle |\phi_{\rm ex}|^2 \rangle 
  = \nu(0) \left(1-\frac{r^2}{R_{\perp}^2}\right)e^{-z^2/\sigma_z^2}\ ,
\end{equation}
with $\nu(0)=2\pi^{-3/2}/(R_{\perp}^2\sigma_z)$.
The terms in the Lagrangian per particle for the above wave function
may be calculated to be
\begin{widetext}
\begin{eqnarray}
  T[\phi_{\rm ex}]
  &=& \frac{\hbar}{2} \int d^3r \left[
  \sum_{i=1}^{N_{\rm v}} \frac{2({\bf r}-{\bf r}_i)\times \dot{{\bf r}}_i}
  {|{\bf r}-{\bf r}_i|^2} - \frac{\dot{\beta} r^2}{d_{\perp}^2}
  - \frac{\dot{\gamma} z^2}{d_z^2} \right]
  |\phi_{\rm ex}|^2\ \nonumber\\
  &\simeq& -\frac{\hbar}{2} \left(\frac{\dot{\beta}R_{\perp}^2}{3d_{\perp}^2}
  + \frac{\dot{\gamma}\sigma_z^2}{2d_z^2}\right)\ ,\label{td3d}\\
  K[\phi_{\rm ex}]
  &=&  \int d^3r \left[\frac{\hbar\omega_{\perp}}{2}
  \left(\frac{1}{\lambda^4}+\beta^2\right)\frac{r^2}{d_{\perp}^2}
  +\frac{\hbar\omega_z}{2}\left(\frac{d_z^4}{\sigma_z^4}+\gamma^2\right)
  \frac{z^2}{d_z^2}
  \right] |\phi_{\rm ex}|^2\ \nonumber\\
  &\simeq& \frac{\hbar\omega_{\perp}}{2}
  \left(\frac{1}{\lambda^4}+\beta^2\right)\frac{R_{\perp}^2}{3d_{\perp}^2}
  + \frac{\hbar\omega_z}{2}\left(\frac{d_z^4}{\sigma_z^4}+\gamma^2\right)
  \frac{\sigma_z^2}{2d_z^2}\ ,
  \label{kin3d}\\
  V[\phi_{\rm ex}] &=& \frac{m}{2} \int d^3r\ 
  (\omega_{\perp}^2 r^2 + \omega_z^2 z^2) |\phi_{\rm ex}|^2\ \nonumber\\
  &\simeq& \frac{\hbar\omega_{\perp}}{2}\frac{R_{\perp}^2}{3d_{\perp}^2}
  + \frac{\hbar\omega_z}{2}\frac{\sigma_z^2}{2d_z^2}\ ,
  \label{pot3d}\\
\mbox{and}\qquad\qquad\qquad&&\nonumber\\
  E_{\rm int}[\phi_{\rm ex}] 
  &=& \frac{g}{2} \int d^3r\ |\phi_{\rm ex}|^4
  \simeq \frac{\sqrt{2}}{3\pi^{3/2}}\frac{bg_{\rm 3D}}{R_{\perp}^2\sigma_z}\ ,
  \label{int3d}
\end{eqnarray}
\end{widetext}
with $g_{\rm 3D}=Ng$.
The angular momentum per particle is
\begin{eqnarray}
  L_z[\phi_{\rm ex}] &=&
  \hbar \int d^3r 
  \left(\frac{r^2}{\lambda^2 d_{\perp}^2}-1\right)
  |\phi_{\rm ex}|^2\nonumber\\
  &\simeq& \hbar\left(\frac{R_{\perp}^2}{3\lambda^2 d_{\perp}^2}-1\right)\ ,
\end{eqnarray}
which is the same as Eq.\ (\ref{lztf}).
Thus the Lagrangian functional can be written as
\begin{widetext}
\begin{eqnarray}
  {\cal L}[\phi_{\rm ex}]
  =-\frac{\hbar}{2} 
  \left(\frac{\dot{\beta}}{3}X^2 + \frac{\dot{\gamma}}{2} Z^2\right)
  - \left[ \frac{\hbar\omega_{\perp}}{2}\left\{\frac{3l^2}{X^2}
  +(\beta^2+1)\frac{X^2}{3}\right\}
  +\frac{\hbar\omega_z}{2}\left\{\frac{1}{2Z^2}
  +(\gamma^2+1)\frac{Z^2}{2}\right\}
  +\frac{4\hbar\omega_{\perp}}{3}\frac{\kappa_{\rm 3D}}{X^2 Z}\right]\ ,
\label{lagrangian3d}
\end{eqnarray}
\end{widetext}
where $X\equiv R_{\perp}/d_{\perp}$, $Z\equiv \sigma_z/d_z$, and
\begin{equation}
  \kappa_{\rm 3D} \equiv \frac{mbg_{\rm 3D}}{2\pi\hbar^2} \frac{1}{\sqrt{2\pi}d_z}
\end{equation}
is the dimensionless interaction strength.

The Euler-Lagrange equations for $\beta$ and $\gamma$ lead to
\begin{eqnarray}
  \dot{X}&=&\omega_{\perp}\beta X\ ,\\
  \dot{Z}&=&\omega_z \gamma Z\ ,
\end{eqnarray}
respectively.
Using these equations, the equations of motion for $X$ and $Z$ yield
\begin{eqnarray}
  &&\frac{1}{\omega_{\perp}^2} X^3\ddot{X} + X^4 - 9l^2 - 
  8\kappa_{\rm 3D}\frac{1}{Z}=0\ ,\\
  &&\frac{1}{\omega_z^2} Z^2\ddot{Z} + Z^3 - \frac{1}{Z} 
  - \frac{8\kappa_{\rm 3D}}{3}\frac{\omega_{\perp}}{\omega_z}\frac{1}{X^2}=0\ .
\end{eqnarray}
The values of $X$ and $Z$ in the equilibrium state, 
$X_0$ and $Z_0$, are given by
the following coupled equations $X_0^4-9l^2-8\kappa_{\rm 3D}Z_0^{-1}=0$ and
$Z_0^3-Z_0^{-1}-(8\kappa_{\rm 3D}/3)(\omega_{\perp}/\omega_z)X_0^{-2}=0$,
and linearized equations for small deviations $\delta X$ and $\delta Z$
from $X_0$ and $Z_0$ are
\begin{equation}
  \frac{1}{\omega_{\perp}^2} X_0^3 \delta\ddot{X} + 4X_0^3 \delta X
  + 8\kappa_{\rm 3D}\frac{\delta Z}{Z_0^2}=0\ ,\label{linear3d1}
\end{equation}
and
\begin{equation}
  \frac{1}{\omega_z^2} Z_0^2 \delta\ddot{Z} + 3Z_0^2 \delta Z 
  + \frac{\delta Z}{Z_0^2} 
  + \frac{16\kappa_{\rm 3D}}{3}\frac{\omega_{\perp}}{\omega_z}\frac{\delta X}{X_0^3}=0\ .\label{linear3d2}
\end{equation}
Now writing $\delta X=A_X e^{i\omega t}$ and $\delta Z=A_Z e^{i\omega t}$,
and using the condition that Eqs.\ (\ref{linear3d1}) and (\ref{linear3d2})
have a non-trivial solution, we finally obtain
\begin{widetext}
\begin{eqnarray}
  \omega^2 &=& \frac{1}{2}
  \left[\left(3+\frac{1}{Z_0^4}\right)\omega_z^2+4\omega_{\perp}^2\right]
  \pm\frac{1}{2}\sqrt{
  \left[\left(3+\frac{1}{Z_0^4}\right)\omega_z^2+4\omega_{\perp}^2\right]^2
  -16\omega_{\perp}^2\omega_z^2 \left(3+\frac{1}{Z_0^4}\right)
  +\frac{512\kappa_{\rm 3D}^2}{3}\frac{\omega_{\perp}^3\omega_z}{X_0^6 Z_0^4}
  }\nonumber\\
  &=& \frac{1}{2}
  \left[\left(3+\frac{1}{Z_0^4}\right)\omega_z^2+4\omega_{\perp}^2\right]
  \pm\frac{1}{2}\sqrt{
  \left[\left(3+\frac{1}{Z_0^4}\right)\omega_z^2+4\omega_{\perp}^2\right]^2
  -16\omega_{\perp}^2\omega_z^2 \left(3+\frac{1}{Z_0^4}\right)
  + 8\omega_{\perp}^2\omega_z^2\left(1-\frac{9l^2}{X_0^4}\right)
  }\ .\qquad
\label{omega3d}
\end{eqnarray}
\end{widetext}
We note that the $1/Z_0^4$ terms have come from the zero-point energy
in the $z$-direction.
From a variational calculation in the rotating frame, we find
$9l^2=X_0^4 \Omega_0^2/\omega_{\perp}^2$, where $\Omega_0$ is the 
angular velocity of the cloud in the equilibrium state for a given
angular momentum.
If one neglects the zero-point energy in the $z$-direction
\cite{note zero-point},
the above expression reduces to
\begin{equation}
 \omega^2 \simeq 2\omega_{\perp}^2 + \frac{3}{2}\omega_z^2
  \pm \frac{1}{2}
  \sqrt{16\omega_{\perp}^4 +9\omega_z^4-16\omega_{\perp}^2\omega_z^2
  -8\omega_z^2\Omega_0^2}\ ,
\label{omega3d_slow}
\end{equation}
which is exactly the same as Eq.\ (45) of Ref.\ \cite{sedrakian} 
(for a polytropic index equal to two) and Eq.\ (10) of Ref.\ \cite{cozzini} 
derived within the hydrodynamic theory in the slow-rotation regime.
This agreement justifies our results
in the present paper obtained with the extended LLL wave function
even for the slow rotation regime.
In the breathing mode, only the coarse-grained density 
and the averaged vortex density are relevant degrees of freedom
when the number of vortices is large;
the local vortex structure, which cannot be described by 
the extended LLL in the slow rotation regime, is irrelevant.
Unlike the ordinary LLL wave function, the extended LLL wave function
can describe the averaged vortex density correctly also in the slow rotation
regime due to the extra degree of freedom $\lambda$,
which is the reason of the above agreement.
(In the slow rotation regime, $T_1\ll T_2$.)

In the rapid rotation limit, where the $z$-dependence of the wave function
corresponds to
the ground state of a particle in a harmonic potential, $Z_0\simeq1$,
and Eq.\ (\ref{omega3d}) leads to
\begin{equation}
  \omega^2\simeq 2\omega_{\perp}^2 + 2\omega_z^2
  \pm 2 \sqrt{(\omega_{\perp}^2-\omega_z^2)^2
  +\frac{1}{2}\omega_z^2(\omega_{\perp}^2-\Omega_0^2)}\ .
\label{omega3d_fast}
\end{equation}
In the limit of $\Omega_0\rightarrow\omega_{\perp}$,
the two frequencies become $\omega=2\omega_{\perp}$ and $\omega=2\omega_z$.
The former value corresponds to the transverse breathing mode
and the latter to the axial one.
Unlike those for the hydrodynamic models \cite{sedrakian,cozzini},
our calculations are therefore able to explain the experimentally obtained
change in the axial breathing mode frequency from the value
$\omega=\sqrt{3}\omega_z$ given by Eq.\ (\ref{omega3d_slow}) to $2\omega_z$
when the interaction energy per particle falls below $\hbar\omega_z$ 
\cite{schweikhard}
(see Ref.\ \cite{axial} for details).


\begin{thebibliography}{99}
\bibitem{note lattice} Here we consider the physical situation
where the vortex lattice extends until $r \sim R$.
Even in the case of the Gaussian density profile, vortices do not exist
in outer region where the mean-field theory breaks down, i.e.,
the number of particle per vortex is of order unity or less:
$1 \agt \pi\ell^2n$ ($\pi\ell^2$ is the area per vortex) or
$r \agt \sigma [\ln{(N\ell^2/\sigma^2)}]^{1/2} 
\sim \sigma[\ln{(N/N_{\rm v})}]^{1/2}$.
\bibitem{note scaling} This can be seen also by the following simple argument.
In two dimensions, since the angular momentum scales as 
$L_z=\int d^2r\ \psi^*({\bf r}\times{\bf p})\psi
\sim \int d^2r\ n (R_{xy}\times mR_{xy}\Omega)\sim m\Omega NR_{xy}^2$,
where $R_{xy}$ is the extent of the cloud
in the $xy$-plane, the rotational kinetic energy scales as
$K_{\rm rot}\equiv\int d^2r\ \psi^*(m/2)(\mbox{\boldmath $\Omega$}
\times {\bf r})^2 \psi \sim m\Omega^2NR_{xy}^2\sim L_z^2/(mNR_{xy}^2)$.
Similarly, the interaction energy in two dimensions is
$E_{\rm int}\equiv (g_{\rm 2D}/2) \int d^2r\ n^2\sim g_{\rm 2D}N^2/R_{xy}^2$,
which has the same $\sim R_{xy}^{-2}$ dependence as $K_{\rm rot}$.
Reflecting the fact that rotation is two-dimensional motion,
the rotational kinetic energy scales as
$K_{\rm rot}\sim L_z^2/(mNR_{xy}^2)$ also in three dimensions
because the vector ${\bf r}$ in $\mbox{\boldmath $\Omega$}\times {\bf r}$
is a transverse vector measured from the rotation axis.
However, in three dimensions,
the interaction energy scales in a different way from $K_{\rm rot}$ as
$E_{\rm int}\equiv (g/2)\int d^3r\ n^2\sim gN^2/(R_{xy}^2 R_z)$,
where $R_z$ is the extent of the cloud in the $z$-direction.
\bibitem{note nonlinear} This statement also holds in large-amplitude
oscillations as will be seen in Section \ref{sect nonlinear}.
However, in that case,
dynamics is determined by the ratio between the radial extent of the cloud
in the equilibrium state, which depends on $\kappa$ and $l$,
and the oscillation amplitude (or the initial velocity of the
radial extent of the cloud driving a breathing oscillation).
Thus the value of $\kappa$ reflects in the time evolution 
of the cloud radius (or the width parameter) in the non-linear oscillation
(not so sensitively because the equilibrium width parameter scales as 
$\sim (l^2+\kappa)^{1/4}$).
\bibitem{note zero-point} Neglecting the zero-point energy in the $z$-direction
is not consistent with the assumption that the axial density profile is 
Gaussian. However, as can be seen from the final result, the detail of the
density profile does not matter for the frequency of the breathing mode.
\end{thebibliography}
\end{document}